\documentclass[fleqn,twoside]{article}
\usepackage{espcrc2}

\usepackage{graphicx}
\usepackage[figuresright]{rotating}

\newcommand{\bea}{\begin{eqnarray}}
\newcommand{\eea}{\end{eqnarray}}
\newcommand{\beq}{\begin{equation}}
\newcommand{\eeq}{\end{equation}}

\newcommand{\AmS}{{\protect\the\textfont2
  A\kern-.1667em\lower.5ex\hbox{M}\kern-.125emS}}
\newcommand{\vev}[1]{\langle #1 \rangle}
\title{$B$ Leptonic Decays and $B-\bar{B}$ Mixing with 2+1 Flavors of Dynamical Quarks}

\author{A. Gray\address{Physics Department, The Ohio State
        University, Columbus, OH 43210, USA.},
        C. Davies\address{Department of Physics \&
               Astronomy, University of Glasgow, Glasgow, G12 8QQ, UK.},
        E. Gulez$^{\rm a}$, 
        G. P. Lepage\address{Laboratory of Elementary Particle Physics,
        Cornell University, Ithaca, NY 14853, USA.}, 
        J. Shigemitsu$^{\rm a}$,
        M. Wingate\address{Institute for Nuclear Theory, University of
        Washington, Seattle, WA 98115, USA.}}
\begin{document}

\begin{abstract}
Calculations of $B$ leptonic decays and $B-\bar{B}$ mixing using NRQCD heavy and Asqtad light valence quarks on the MILC dynamical configurations are described. Smearing has been implemented to substantially reduce the statistical errors of the matrix elements needed for the determination of $f_B$. The four-fermion matrix elements needed for the determination of $f_{B_s}^2B_{B_s}$ have been calculated and a preliminary result is given.
\vspace{1pc}
\end{abstract}

\maketitle

\section{Introduction}
$B-\bar{B}$ mixing is a key process in standard CKM analysis. 
Calculations of the $B(B_s)$ leptonic decay constant $f_B$($f_{B_s}$) and bag parameter $B_B$($B_{B_s}$) are needed to constrain $V_{td}$($V_{ts}$).  In particular, we aim to calculate the ratio $f_{B_s}\sqrt{B_{B_s}}/f_B\sqrt{B_B}$ accurate to the few percent level allowing a determination of $V_{ts}/V_{td}$ in which theoretical errors do not dominate those from experiment.

The Asqtad improved staggered formulation has allowed precise determinations of a variety of quantities \cite{Davies:2003ik}, and recent progress has allowed the use of improved staggered valence quarks in heavy-light simulations \cite{Wingate:2002fh}. 
Here we report on the progress of $B$ leptonic decay and mixing calculations using  NRQCD heavy quarks and improved staggered (Asqtad) light quarks. We use the  MILC $n_f=2+1$ dynamical configurations \cite{Bernard:2001av}.

\section {$B$ leptonic decays}
In previous work \cite{Wingate:2003ni} the chiral behavior of $f_{B_s}\sqrt{m_{B_s}}/f_{B}\sqrt{m_{B}}$ was masked by large statistical errors. Here these errors have been substantially reduced with the introduction of smearing. Two dynamical ensembles were used; one incorporating a light dynamical quark mass $m_f\equiv m_{u,d}=m_s/4$ while the other has $m_f=m_s/2$ where $m_s$ is the physical strange quark mass. The valence light quark mass $m_q$ was varied from $m_s$ to $m_s/8$, i.e. some runs were in the partially quenched approximation. For $b$ quarks, the standard tadpole improved Lattice NRQCD action correct through ${\cal O}(1/{(am_b)^2})$ was used along with zeroth and first order in $1/{am_b}$ currents. $a^{-1}$ and $m_b$ were fixed by $\Upsilon$ while $m_{u,d}$ and $m_s$ were fixed by $\pi$ and $K$ respectively. 

A ground state hydrogenic style wavefuction was used to smear the heavy quark at both source and sink and the optimal radius was found as that which minimized the fit errors while maintaining a reasonable $\chi^2/\mbox{dof}$.
For each combination of source and sink smearing, the functional form of the $k^{th}$ order in $1/am_b$  correlator is given by the oscillating function
\beq
G(t)=\sum_{j=0}^{n_{exp}-1} C^{(kj)}(-1)^{jt}e^{-m_jt}.
\label{}
\eeq
Bayesian multi-exponential fits were done to the $k=0,1$ correlators, and 
$\Phi_B^{(k)}$ were determined as 
\beq                 
\Phi_B^{(k)}=\sqrt{2C^{(k0)}}.                                
\label{}                                                                      
\eeq  
$f_B$ correct through 1-loop is given through
\bea
\Phi_B\equiv f_B\sqrt{m_B}&=&(1+\alpha_s\rho_0)\Phi_B^{(0)}\nonumber \\
&+&(1+\alpha_s(\rho_1+\rho_2))\Phi^{(1)}_{B,sub}
\label{}
\eea
where
\beq
\Phi^{(1)}_{B,sub}=\Phi_B^{(1)}-\alpha_s\zeta\Phi_B^{(0)}
\label{}
\eeq
and the factors $\rho_0,\rho_1,\rho_2,\zeta$ have been perturbatively calculated \cite{gulez:2004}.

Figure 1 compares $\Phi_B^{(0)}$ values
 from single correlator fits to simultaneous multi-correlation matrix (smeared source and sink) and vector (smeared source only) fits.
  It can be seen that the errors are significantly reduced in the matrix case.

Figure \ref{fig:ratxx} shows the effects of smearing on $\xi_{\Phi}={\Phi_{B_s}}/{\Phi_{B}}$. Note that the most chiral partially quenched non-smeared result incorporates only half the statistics of its smeared counterpart, while the statistics match in the other cases. Fully unquenched and partially quenched chiral extrapolations which incorporate staggered taste breaking effects \cite{bernardstag} are underway to the new smeared data, in addition to more fully unquenched simulations at $m_q/m_s=1/8$.

\begin{figure}[t] 
\includegraphics[width=5.7cm,angle=-90]{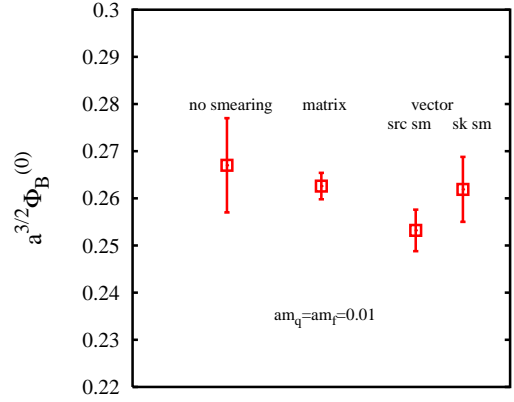}
\label{fig:v010}
\caption{Determinations of $a^{3/2}\Phi_B^{(0)}$ using single correlator (no smearing), matrix (source and sink smearing) and vector (source or sink smearing) fits.}
\end{figure}

\begin{figure}[t] 
\includegraphics[width=5.9cm,angle=-90]{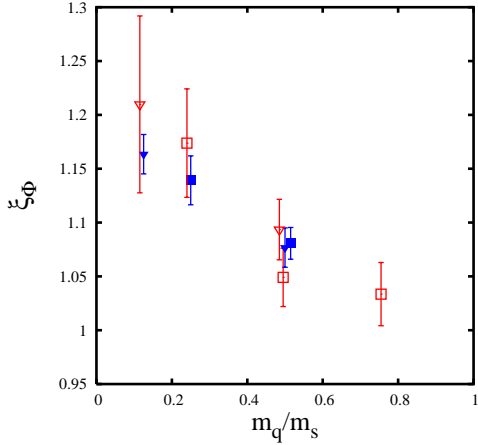}
\label{fig:ratxx}
\caption{$\xi_{\Phi}$ vs. ratio of light and strange valence quark masses. Closed symbols indicate that smearing has been used while open symbols indicate smearing has not been used. Squares represent fully unquenched results and triangles represent results in the partially quenched approximation.
}
\end{figure}

\section {$B-\bar{B}$ Mixing}
3-point correlators were generated with the creation of a  $\bar{B}$ at $t=-t_{\bar{B}}$, conversion to a $B$ with the use of a four-fermion operator ${\cal O}$ at $t=0$ and destruction of the $B$ at $t=t_B$. The same simulation parameters as in the $B$ leptonic decay case were used but only so far with $m_f=0.01$, $m_q=0.04$ (i.e the $B_s$). No smearing has yet been performed, and so far only zeroth order in $1/am_b$ currents have been considered.

The continuum four fermion operator $\vev{{\cal O}_L}^{\overline{MS}}$ is given in terms of lattice operators at 1-loop by
\bea
a^6 \vev{{\cal O}_L}^{\overline{MS}} &=& [1+ \rho_{LL} \, \alpha_s] \; \vev{{\cal O}_L}_{lat}\quad \nonumber \\
&+& \quad \rho_{LS} \, \alpha_s \; \vev{{\cal O}_S}_{lat}\label{}\eea
where
\bea
 {\cal O}_L &=& [\, \bar{\psi}_Q \, \gamma^\mu (1-\gamma_5) \, \psi_q\,] \;
[ \, \bar{\psi}_{\bar{Q}}\, \gamma_\mu (1-\gamma_5) \, \psi_q \,] \nonumber \\
{\cal O}_S  &=& [\, \bar{\psi}_Q\,(1-\gamma_5) \, \psi_q\,] \;
[ \, \bar{\psi}_{\bar{Q}}\,  (1-\gamma_5) \, \psi_q \,]
\label{}
\eea
and $\rho_{LL}$ and $\rho_{LS}$ have been calculated perturbatively \cite{gulez:2004}.

\begin{figure}[t] 
\includegraphics[height=7.5cm,angle=-90]{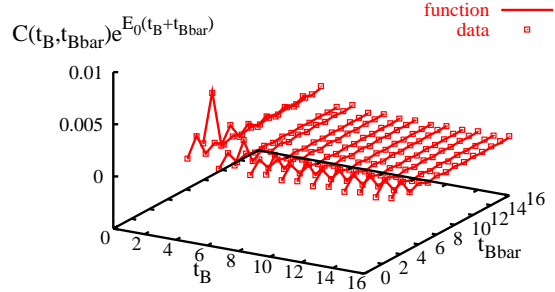}
\caption{Fit result to four-fermion correlator with ground state energy contributions removed. Squares show data points while lines show fit function. }
\label{fig:BBbar}
\end{figure}

The lattice correlators have the functional form
\bea
C(t_B,t_{\bar{B}})&=&\sum_{j,k=0}^{n_{exp}-1} A_{jk}(-1)^{jt_B}e^{-m_jt_B}\nonumber \\ &\mbox{ }&\mbox{ }\mbox{ }\mbox{ }\mbox{ }\mbox{ }\mbox{ }\mbox{ }\mbox{ }\mbox{ }\mbox{ }*(-1)^{kt_{\bar{B}}}e^{-m_kt_{\bar{B}}}.
\label{}
\eea
A preliminary Bayesian fit has been successful for $n_{exp}=4$, as can be seen in Figure \ref{fig:BBbar} which compares the data and fit function of  the effective amplitude (correlator with ground energy contributions removed). The oscillations in both $t_B$ and $t_{\bar{B}}$ can clearly be seen to be nicely represented by the fit function, and the fit has a good $\chi^2/\mbox{dof}$. Work is underway to obtain fits at other $n_{exp}$ in order to solidify these results.

$\vev{{\cal O}_{L,S}}_{lat}$ are determined through 
\beq
\frac{1}{2M_Ba^3} \; \vev{{\cal O}_{L,S}}_{lat}=
\frac{A_{00}^{({\cal O}L,S)}}{\xi_{BB}}                         
\label{}
\eeq 
where $\xi_{BB}=C^{(00)}$; the 2-point ground state leading order in $1/am_b$ amplitude from the previous section.
Note that the fit is done directly to the 3-point correlators without first taking the ratio over the 2-point. This means that fits at low t, where the 
correlator has good statistical noise, are possible using multiple 
exponentials without the inefficiency and ambiguity of needing a
plateau.

$B_B$ is defined through                                                 
\beq
\vev{{\cal O}_L}^{\overline{MS}} = \frac{8}{3} f_B^2 M_B^2 B_B
\label{}
\eeq
and our results preliminarily give 
\beq
f_{B_s} \, \sqrt{B_{B_s}(m_b)} = 0.197(16)(28)\mbox{GeV}. 
\label{}
\eeq 
The first error arises from the fact that the fits are still preliminary and the second error is from systematics, the main contributions being from the neglection of ${\cal O}(\Lambda_{QCD}/m_b)$ and ${\cal O}(\alpha_s^2)$ contributions.

\section{Conclusions}
We report here on the status of $B$ leptonic decays and $B-\bar{B}$ mixing simulations with NRQCD heavy and Asqtad light quarks. The use of smearing has significantly reduced the statistical errors for the leptonic decay case. Chiral extrapolations and more fully unquenched simulations are in progress. A successful fit to the $B\bar{B}$ correlator has allowed a preliminary result for  $f_{B_s} \, \sqrt{B_{B_s}}$. 
\section{Acknowledgments}
 This work was supported by the DOE, PPARC and NSF. Simulations were carried out at NERSC.

\end{document}